\documentclass{ws-procs9x6}

\begin{document}

\title{Isobar rescattering model and light scalar mesons}

\author{ J. SCHECHTER}

\address{ Physics Department, \\
Syracuse University, \\
Syracuse, NY 13244-1130, USA. \\
E-mail: schechte@phy.syr.edu}

\maketitle

\abstracts{
We use a toy model to discuss the problem of
parameterizing the possible contribution of a light scalar meson,
sigma, to the final state interactions in the non leptonic decays
of heavy mesons.}


\vspace{0.1in}

\section{Introduction}

\begin{figure}[htbp]
\centering
{\includegraphics[width=10.00cm,height=7.00cm,clip=true]
{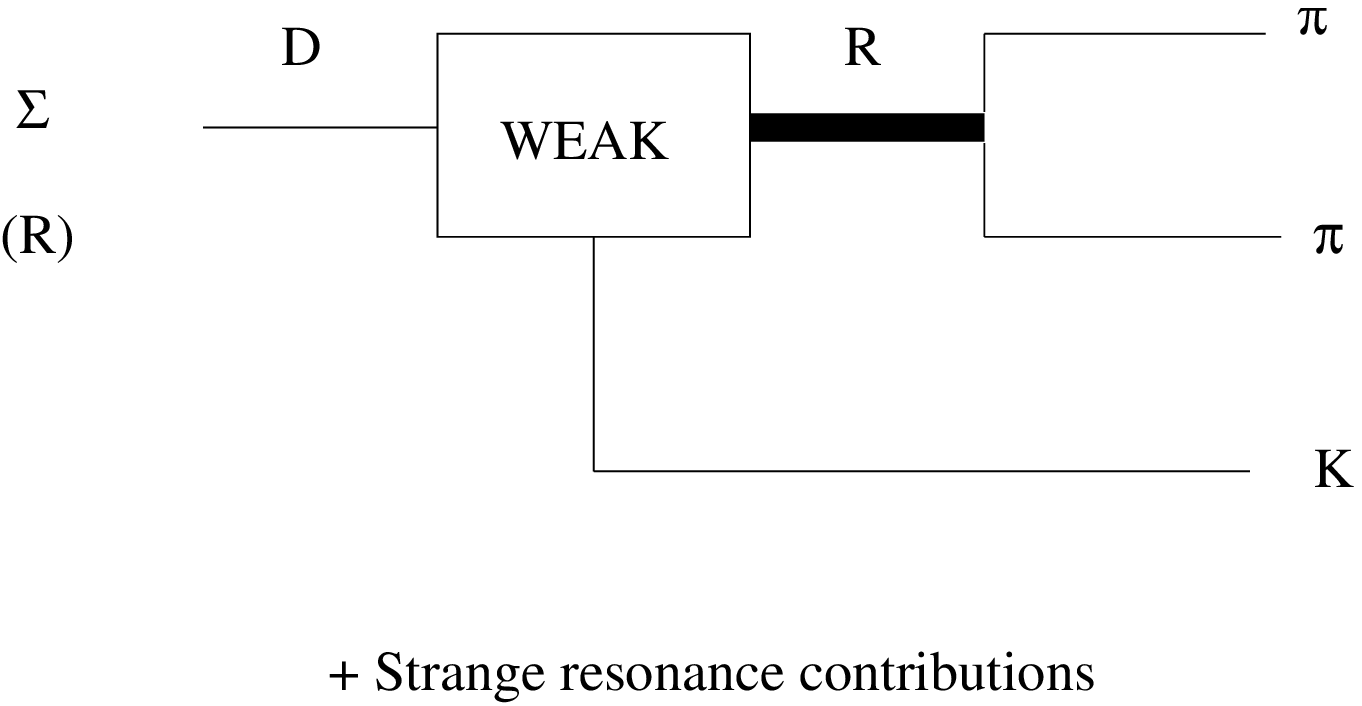}}
\caption[]{Isobar model for the decay D $\rightarrow \pi\pi{\bar K}$}
\label{decay}
\end{figure}

    A nice feature of recent results from BABAR, BELLE, CESAR
and FERMILAB is the 
historically unprecedented number of points in the Dalitz diagrams
obtained for decays like $D^0 \rightarrow \pi^+ \pi^- {\bar K_0}$.
It is not only favorable for learning more about CP violation
and associated problems in the weak interactions but
for learning about properties of resonances in the strong interactions.
This involves a resonance dominated description \cite{asner} (so-called isobar model)
as illustrated in Fig. \ref{decay}. Typically, a Breit-Wigner shape
as well as an arbitrary overall complex number is assumed
for each resonance. While it is hard to rigorously justify
such a procedure from fundamental QCD, it is made plausible
by the large $N_C$ approximation \cite{largeN} which actually suggests
that such a procedure should hold when the resonances are quark anti-quark
composites. Of course, this kind of procedure has worked very well
in the past for many established resonances like the
$\rho$(770).

    Recently, there has been a great deal of interest
(See ref\cite{E791}-ref\cite{oller}) in the possibility
of observing the seemingly elusive light scalar mesons, like the sigma ($f_0$(600))
and its strange analog, kappa, from such Dalitz diagrams. Evidence for
their existence had been previously obtained by many workers based on constructing
theoretical models and comparing with experimentally obtained phase shifts.
This evidence suggests that these resonances are ``hidden"- they do not have obvious
Breit Wigner shapes but seem to contain a lot of interference with the
background. Furthermore, their structure seems more likely to be 
two quarks- two antiquarks rather than quark- antiquark; this
is deduced from the fact that, when amalgamated in a putative SU(3) nonet,
they display an upside-down ordering with the degenerate
iso-singlet, isotriplet states ($f_0$(980), $a_0$(980)) appearing as heaviest
rather than lightest members. It should be noted that such a ``four quark"
structure is theoretically disfavored for resonances in the leading tree order
of the large $N_c$ approximation. So, on both experimental and theoretical
grounds, one may expect that the parameterization of these possible resonances
in the Dalitz analysis might very well be different from the parameterization
of ordinary quark- antiquark resonances.

    What is the crucial ingredient in a suitable parameterization of low
energy scalar resonances? We will argue that it is chiral symmetry.

\section{Chiral symmetry}
 
    The problem of the scalar channel in pion physics goes back about fifty years.
Using the Yukawa (pion exchange) model, the value of the pi-nucleon coupling constant,
$g_{\pi NN}$ was well established from the long distance behavior of the nucleon-
nucleon potential. However, when this value of coupling constant was applied
to try to calculate the s-wave pion nucleon scattering length using the nucleon 
exchange diagrams, the answer
turned out to be more than ten times larger than experiment. The cure was eventually
realized to involve upgrading the iso-spin or SU(2) symmetry to two 
 separate iso-spin symmetries for the left and right handed components of the nucleon. This 
chiral SU(2) symmetry would be manifest if the nucleons were massless. However, Nambu
had the insight to propose that the symmetry is there anyway and that the nucleon receives a 
mass because the symmetry is ``spontaneously broken". The simplest
realization of this mechanism is embodied in the Gell Mann - Levy linear sigma model \cite{gl}.
The chiral symmetry meson multiplet in this model contains, in addition to the pions,
a scalar particle denoted the sigma. When the mass of the sigma is considerably
higher than the pion mass, the effect of sigma exchange cancels out almost all
of the much too large result for the s-wave pion nucleon scattering and leaves
the right answer.

    The same situation prevails for the s-wave pion pion scattering amplitude.
The meson part of the chiral invariant linear sigma model Lagrangian is given in
terms of the pion and sigma fields as:
\begin{equation}
L = -\frac{1}{2}(\partial_\mu\pi
\cdot\partial_\mu\pi +\partial_\mu \sigma
\partial_\mu \sigma)
+a(\sigma^2+\pi^2)
-b(\sigma^2+\pi^2)^2 ,
\label{LsMLag}
\end{equation}
where the real parameters $a$ and $b$ are both taken positive to
insure spontaneous breakdown of chiral symmetry.  The vacuum value
$\left< \sigma \right>$ of the $\sigma$ field is related to the pion
decay constant as
\begin{equation}
{\rm F}_\pi = \sqrt{2} \left< \sigma \right>,
\label{Fpi}
\end{equation}
where ${\rm F}_\pi = 0.131$ GeV.
The parameter $a$ is given by
\begin{equation}
a = \frac{1}{4} m_{\sigma B}^2,
\end{equation}
where $m_{\sigma B}$ is the tree level (or bare) value of the sigma mass.
The only unknown parameter at the moment in this simple model is
$m_{\sigma B}$ since one has $b=m_{\sigma B}^2/(2F_{\pi})^2$. Then,
 calculating the conventional pion pion amplitude as the sum of
the 4-point pion term plus the sigma exchange term 
gives the result,
\begin{equation}
A(s, t, u)=\frac{2(m_{\sigma B}^2-m_\pi^2)}{F_\pi^2}[(1-
\frac{s-m_\pi^2}{m_{\sigma B}^2-m_\pi^2})^{-1} -1].
\label{pipiamp}
\end{equation}
Here, $s,t,u$ are the Mandelstam variables and we have included the
effect of the small pion mass by adding a term linear in the field $\sigma$
to Eq. (\ref{LsMLag}).

    The last entry ``-1" in this equation represents the 4 point pion term. Near
threshold ($s=4m_\pi^2$) and for $m_{\sigma B}^2$ large compared to $m_\pi^2$, most of this is 
seen to be canceled away and one is left with Weinberg's formula \cite{w} for near threshold scattering,
\begin{equation}
A(s, t, u) = \frac{2}{F_\pi^2}(s-m_\pi^2),
\label{pipithresh}
\end{equation}
which works very well. The inelegant feature of the present derivation is that
it arises as the near cancellation between two large quantities. This can be cured by
``integrating out" the  relatively heavy sigma (as was done for a different reason already
 by Gell Mann and Levy). Then the pion field appears non-linearly; this non-linear sigma
model treats the interaction strength as properly ``small" and is convenient to use
for further calculations (Chiral perturbation theory program). However, we shall use the original
 linear model here since our purpose is to provide a simple toy model to illustrate how to
parameterize the effects of the
sigma meson. From the above we can already draw the lesson that a simple pole term for the sigma
without the associated 4 pion contact term is likely to be misleading. Both are needed
 for the chiral symmetry to hold.

    It is amusing that the SU(2) linear sigma model is identical to the Higgs 
sector of the standard electroweak theory when one scales up $F_\pi$ by a factor of
about 2660. The sigma becomes the Higgs particle and the pions become the 
longitudinal components of the W and Z bosons. Our present understanding of
``precision" electroweak theory suggests that the quantity $m_{\sigma B}$
scales up by a slightly reduced factor so that the dimensionless quantity
b (which measures the strength of the Higgs self coupling) is smaller
than for the possible low energy QCD application.

\section{Pion pion scattering}

The needed I=J=0 partial wave amplitude for pi pi scattering at tree level is
obtained from Eq. (\ref{pipiamp}) as:
\begin{equation}
\left[T^{0}_{0}\right]_{\rm tree}(s) =  \alpha \left( {s}
\right) + \frac{\beta (s)}{m_{\sigma B}^2 - s}
\label{partialpipiampl}
\end{equation}
where
\begin{eqnarray}
\alpha \left( \rm{s}\right) &=&  \frac { \sqrt {1 -
\frac{4m_\pi^2}{s}}}{32\pi F_\pi^2} \left({m_{\sigma B}^2 } -
{m_\pi}^2 \right)
\left[ -10 + 4 \frac{{m_{\sigma B}^2} - {m_\pi}^2}{s - 4 {m_\pi}^2} \rm{ln}
\left( \frac{{m_{\sigma B}^2} + s - 4{m_\pi}^2}{{m_{\sigma B}^2}} \right) \right],  \nonumber \\
\beta (s) &=& \frac {3 \sqrt {1 -
\frac{4m_\pi^2}{s}}}{16 \pi F_\pi^2}   {\left( {m_{\sigma B}^2} - {m_\pi}^2 \right)}^2.
\label{alphabeta}
\end{eqnarray}
The normalization of the amplitude $T_0^0(s)$ is
given by its relation to the partial wave S-matrix
\begin{equation}
S_0^0(s) = 1 + 2iT_0^0(s).
\label{Smatrix}
\end{equation}

 While, as just discussed, this tree-level formula
works well at threshold it does involve large coupling constants and
cannot be expected to be a priori reasonable even several hundred MeV
above threshold.  In addition, at the point $s=m_{\sigma B}^2$, the
amplitude Eq. (\ref{partialpipiampl}) diverges. The solution to this
problem, which is adopted in the 
conventional isobar rescattering parameterizations,
 is to include a phenomenological width term in the denominator
by making the replacement:
\begin{equation}
\frac{1} {m_{\sigma B}^2  - s} \longrightarrow \frac{1}
{m_{\sigma B}^2  - s - i m_{\sigma B}\Gamma }.
\label{conventionalreg}
\end{equation}
However this standard approach is not a good idea in the present
case.  As emphasized by Achasov and Shestakov \cite{AS94},
the replacement Eq. (\ref{conventionalreg}) completely destroys the
good threshold result
which is a consequence of chiral symmetry.  This is readily
 understandable since the
threshold result was seen to arise from a nearly complete
cancellation between the first and second terms of
Eq. (\ref{partialpipiampl}).   However, the pole in the linear sigma
model can be successfully handled by using, instead of
Eq. (\ref{conventionalreg}), K-matrix regularization, which instructs
us to adopt the exactly unitary form
\begin{equation}
S_0^0 (s) = \frac {1 + i \left[{T^{0}_{0}}\right]_{\rm tree}(s)  }{1 -
i \left[{T^{0}_{0}}\right]_{\rm tree}(s)  }
\label{regularization}
\end{equation}
Using Eq. (\ref{Smatrix}) we get
\begin{equation}
T_0^0(s) = \frac { \left[{T^{0}_{0}}\right]_{tree}(s)}{ 1 - i
\left[{T^{0}_{0}}\right]_{tree}(s)}.
\label{Tregularization}
\end{equation}
Near threshold, where $\left[{T^{0}_{0}}\right]_{tree}(s)$ is small,
this reduces to $\left[{T^{0}_{0}}\right]_{tree}(s)$ as desired.
Elsewhere it provides a unitarization of the theory which is seen to
have the general structure of a ``bubble-sum''.  We will adopt this
amplitude as our toy model for (the strongly
coupled) QCD in the low energy I=J=0 channel.
    
\begin{figure}[htbp]
\centering
{\includegraphics[height=10cm,angle=270]{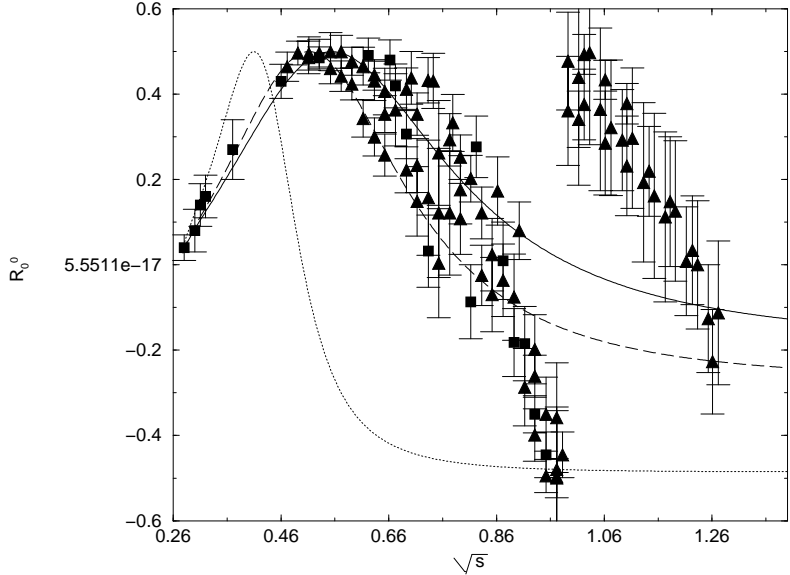}}
\caption[]
{Comparison with experiment of Real part of the I=J=0 $\pi \pi$ scattering
amplitude in the SU(2) Linear Sigma Model, for $m_{\sigma B} =0.5$ GeV (dots),
$m_{\sigma B} =0.8$ GeV (dashes) and $m_{\sigma B}=1$
GeV (solid).  Experimental data \cite{pipidata} are extracted from Alekseeva
{\it et al} (squares) and Grayer {\it et al} (triangles).}
\label{finalSU2LsMRamp}
\end{figure}

      The obvious question is whether this toy model can explain the
experimental data. The only parameter available is $m_{\sigma B}$.
In Fig. \ref{finalSU2LsMRamp} the real part of the $T^0_0$ amplitude
(sufficient in the elastic regime) is plotted against existing data
for several values of $m_{\sigma B}$. It is seen that there is at least
a rough
fit to the data up till about 0.8 GeV if $m_{\sigma B}$ lies in the range
0.8 - 1.0 GeV. Clearly, the energy region
between about 0.8 and 1.2 GeV is not at all fit by the model.  However
this is due to the neglect of a second scalar resonance which is
expected to exist in low energy QCD.  As shown in
\cite{BFMNS01} if the SU(2) Lagrangian is ``upgraded''
to the three-flavor case (so that another scalar field $\sigma^\prime$
identifiable with the $f_0(980)$ is contained) the entire region
 up to about $\sqrt{s} = 1.2$ GeV can
be reasonably fit with the same K matrix unitarization scheme.  

   In assessing the validity of this toy model for low energy QCD
one should also consider the role of the vector mesons.  These are
known to be important in many low energy processes and give
the dominant contributions to the ``low energy constants'' of the
chiral perturbation theory expansion.  Nevertheless it was found
\cite{HSS2} that, while rho meson
 exchange does make a contribution to low energy s-wave pion pion scattering,
 its inclusion
does not qualitatively change the properties of the light $\sigma$
resonance which seems crucial to explain the I=J=0 partial wave.  More
specifically, the effect of the rho meson raises the $\sigma$ mass by
about 100 MeV and lowers its width somewhat.

\begin{figure}[htbp]
\centering
{\includegraphics[height=9cm,angle=0]{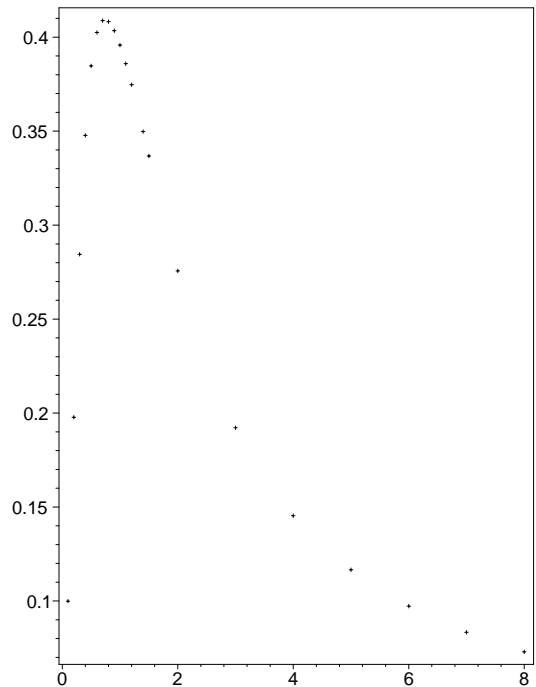}}
\caption
{Plot of $\sqrt {{\rm Re} (z_0)}$ in GeV as a function of $m_{\sigma B}$
in GeV. Here we have set
$m_\pi = 0$.   }
\label{LsMpoleSqrtRez0}
\end{figure}

     Thus, we have seen that the given toy model provides a reasonable
description of low energy s-wave pion pion scattering. At the same time we have seen
a couple of its drawbacks and how to improve them. Before going to the
rescattering application, it seems worthwhile to mention an interesting feature
of the K-matrix unitarization.
The criterion for the model to be in the non-perturbative region is
that the dimensionless coupling constant 
$b = { \left( \frac{m_{\sigma B}}{2F_\pi} \right) }^2$
is greater than unity. Here $b$ is around 10.
Thus one might
expect the physical parameters like the sigma mass and width to differ
from their ``bare'' or tree-level values.  To study this we look at
the complex sigma pole position in the partial wave amplitude in
Eqs. (\ref{Tregularization}) and (\ref{alphabeta}):

\begin{equation}
T_0^0(s) = \frac { (m_{\sigma B}^2 - s) \alpha (s) + \beta(s) }
{ (m_{\sigma B}^2 - s)[1 - i \alpha(s)] - i \beta(s)}.
\label{Tunitary}
\end{equation}
This is regarded as a function of the complex variable $z$ which
agrees with $s + i \epsilon$ in the physical limit.  The pole position
$z_0$ is then given as the solution of:

\begin{equation}
(m_{\sigma B}^2 - z_0)[ 1 - i \alpha(z_0)] - i \beta(z_0) = 0.
\label{poleeq}
\end{equation}
Note that $\alpha(s)$ remains finite as $q^2=s-4m_\pi^2 \rightarrow
0$, so there are no poles due to the numerator of Eq. (\ref{Tunitary}).
In Fig. \ref{LsMpoleSqrtRez0} we show how $\sqrt{Re(z_0)}$, a measure
of the ``physical" sigma mass, depends on the choice of $m_{\sigma B}$.
Note that there is a maximum physical mass and correspondingly two
different values of  $m_{\sigma B}$ will result in the same physical mass.
In fact, the best fit value lies a bit to the right of the peak; there the physical
mass is considerably less than the bare mass. Similarly, the physical width
 comes out considerably smaller than the bare width.

It is amusing to observe that the peak of
$|T^0_0(s)|$ occurs at the bare sigma mass, $s=m_{\sigma B}^2$. This
may be seen by noting that, since $T^0_0(s)$ can be expressed in
terms of the phase shift as $\exp(i\delta^0_0(s))\sin\delta^0_0(s)$,
the peak will occur where $|\delta^0_0(s)|=\pi/2$.
In other words, the peak will occur where $T^0_0(s)$ is pure
imaginary. This is immediately
 seen from Eq (\ref{Tunitary}) to be the case when $s=m_{\sigma B}^2$.
It should also be remarked that the shape of $|T^0_0(s)|$ differs
considerably from a corresponding Breit-Wigner shape for the 
parameter choice which fits the s-wave pion pion scattering data.

\section{Rescattering}

    We have argued against using for the rescattering term representing the sigma
in Fig. \ref{decay},
the usual choice:
\begin{equation}
\frac{g_{\sigma\pi\pi}}{m_{\sigma B}^2  - s - i m_{\sigma B}\Gamma },
\label{notpreferred}
\end{equation}
where $g_{\sigma\pi\pi}$ is the sigma pi pi coupling constant in the model (which
will anyway get absorbed in an unspecified ``production" factor). In the K-matrix
unitarization we do not introduce the width by hand. Rather the rescattering is
represented by the ``bubble sum" factor $(1-i[T^0_0]_{tree})^{-1}$
 as in Eq. (\ref{Tregularization}). Then, the proper rescattering factor 
 \cite{AS94} is,
\begin{equation}
\frac{  g_{\sigma \pi \pi} }{m_{\sigma B}^2 - s} \frac{ 1 }
{ 1 - i{[T_0^0]}_{\rm tree} } 
= \frac{  g_{\sigma \pi \pi} }{m_{\sigma B}^2 - s} \cos \delta_0^0
\exp{i \delta_0^0},
\label{FSI}
\end{equation}
where we used the relation for the phase
shift $\delta_0^0$ in the K-matrix unitarization scheme:  $\tan
\delta_0^0 = {[T_0^0]}_{\rm tree}$, where in turn the unitarized I=J=0 S-matrix
element is given by $S_0^0 = \exp{2i\delta_0^0}$.
It is important to examine the quantity $\cos \delta_0^0
(s)$.  Using
Eq. (\ref{partialpipiampl}) it is straightforward to find
\begin{equation}
\cos \delta_0^0(s) = \frac { m_{\sigma B}^2 - s}{[ 
( m_{\sigma B}^2 - s)^2 + ( \alpha(s)
( m_{\sigma B}^2 - s) + \beta(s))^2 ]^{1/2}}.
\label{cosdelta}
\end{equation}
The numerator cancels the pole at $s = m_{\sigma B}^2$ in
 Eq. (\ref{FSI}) so one has a finite result
for the rescattering factor expressed in terms
of the functions $\alpha(s)$ and $\beta(s)$ defined in
 Eq.(\ref{alphabeta}). In contrast to $|T_0^0|$, the magnitude of the
rescattering factor $|\cos \delta_0^0e^{i\delta_0^0}/( m_{\sigma B}^2 - s)|$  peaks 
at the much smaller physical value rather than the bare value of the sigma mass \cite{abfns}.
This is understandable since there is a trivial numerator and the denominator
 has the same structure as the denominator used to find the physical pole position.

    To sum up, we have stressed that it is necessary to take 
chiral symmetry into account when parameterizing the final state rescattering into
two pions in the I=J=0 channel. We presented a simple toy model \cite{note} which
seems able to do the job. In this model the quantity $m_{\sigma B}$ may be considered a parameter.
Evidently there is a lot more to do in this area and hopefully the possibility of
improving our understanding of low energy strong dynamics will emerge.

\section*{Acknowledgments}
   This talk is mainly based on reinterpreting \cite{abfns} which deals with
production of longitudinal electroweak gauge bosons by the "gluon fusion" mechanism.
More complete referencing is included there.
I am happy to thank my collaborators A. Abdel-Rehim, D. Black, A. H. Fariborz, M. Harada,
R. Jora, S. Moussa, S. Nasri and F. Sannino for many exciting discussions.
Amir Fariborz deserves a vote of thanks for excellently organizing this
stimulating conference. An earlier version
of the talk was
presented at the BaBar Dalitz Workshop (SLAC, Dec. 5, 2004) and I am grateful to B. Meadows and 
A. Palano for inviting me to that interesting event.
The talk was written up at the University of Valencia; I
sincerely thank J. W. F. Valle and his colleagues there
for providing gracious hospitality and a stimulating physics 
environment.
 The work was supported in part  
by the U. S. DOE under Contract no. DE-FG-02-85ER 40231.


\begin{thebibliography}{0}

\bibitem{asner}See, for example, D. M. Asner et al, arXiv:hep-ex/0311033.

\bibitem{largeN}G.'t Hooft, Nucl. Phys. {\bf B72}, 461 (1974); E. Witten, 
Nucl. Phys. {\bf 160}, 57 (1979).

\bibitem{E791}E791 Collaboration, Phys. Rev. Lett. {\bf 89}, 121801 (2002).

\bibitem{CLEO}CLEO Collaboration, Phys. Rev. Lett. {\bf 89}, 251802 (2002).

\bibitem{Belle}Belle Collaboration, arXiv:hep-ex/0308043.

\bibitem{BaBar}BaBar Collaboration, arXiv:hep-ex/0408088.

\bibitem{oller}J. A. Oller, arXiv:hep-ph/0411105.

\bibitem{gl}M. Gell-Mann and M. Levy, Nuovo Cimento {16}, 705 (1960). 

\bibitem{w}S. Weinberg Phys. Rev. Lett. {\bf 18}, 188 (1967).

\bibitem{AS94} N.N. Achasov and G.N. Shestakov, Phys. Rev. {\bf
D49}, 5779 (1994).

\bibitem{BFMNS01}D. Black, A.H. Fariborz, S. Moussa, S. Nasri and
J. Schechter, Phys. Rev. D {\bf 64}, 014031 (2001).

\bibitem{HSS2}M. Harada, F. Sannino and J. Schechter, Phys. Rev. Lett. {\bf 78}, 1603 (1997).

\bibitem{pipidata}E.A. Alekseeva {\it et al}., Sov. Phys. JETP {\bf
55}, 591 (1982), G. Grayer {\it et al}., Nucl. Phys. {\bf B75}, 189 (1974).

\bibitem{abfns}A. Abdel-Rehim, D. Black, A. H. Fariborz, S. Nasri and J. Schechter,
Phys. Rev. D {\bf 68}, 013008 (2003).

\bibitem{note}We argued that the sigma meson is likely to be an exotic
 resonance with a four quark structure, which does not appear as a pole
 at leading order in the $1/N_c$ approximation. However, in the linear sigma
model it is the chiral partner of the pion and hence presumably a p wave
quark antiquark state. As far as the $1/N_c$ expansion is concerned, it should
then appear at leading order. We can explain this feature by saying that
the bare sigma in the model is heavy and as such not apparently relevant for 
dynamics below 1 GeV. However, the unitarization, which has the effect of
emphasizing the two meson  (or four quark) component, is a sub leading
large $N_c$ feature which is important in lowering the bare sigma mass in our
world of $N_c$ =3. In addition there is another effect which seems to push in
the same direction. It has been argued that in the linear sigma model approach
 it is better to start with two chiral fields- one representing bare two quark states
and  the other bare four quark states. The low lying scalars in this picture
tend to have relatively large four quark components. See section V of \cite{BFMNS01}
above, M. Napsuciale and S. Rodriguez, arXiv:hep-ph/0407037 and A. H. 
Fariborz, 
R. Jora and J. Schechter, arXiv:hep-ph/0506170.
\end{thebibliography}
\end{document}